# The Habitable Zone Planet Finder: A Proposed High Resolution NIR Spectrograph for the Hobby Eberly Telescope to Discover Low Mass Exoplanets around M Dwarfs

Suvrath Mahadevan*[a,b], Larry Ramsey[a,b], Jason Wright[a,b], Michael Endl[c], Stephen Redman[a], Chad Bender[a,b], Arpita Roy[a,b], Stephanie Zonak[a], Nathaniel Troupe[d], Leland Engel[d], Steinn Sigurdsson[a,b,f], Alex Wolszczan[a,b], Bo Zhao[e]

[a]Department of Astronomy & Astrophysics, Pennsylvania State University, 525 Davey Lab, University Park, PA-16802; [b]Center for Exoplanets & Habitable Worlds, Penn State; [c]McDonald Observatory, University of Texas, Austin; [d]Department of Mechanical Engineering, Penn State; [e]Department of Astronomy, University of Florida, [f]Penn State Astrobiology Research Center, Penn State

## ABSTRACT

The Habitable Zone Planet Finder (HZPF) is a proposed instrument for the 10m class Hobby Eberly telescope that will be capable of discovering low mass planets around M dwarfs. HZPF will be fiber-fed, provide a spectral resolution R~ 50,000 and cover the wavelength range 0.9-1.65μm, the Y, J and H NIR bands where most of the flux is emitted by mid-late type M stars, and where most of the radial velocity information is concentrated. Enclosed in a chilled vacuum vessel with active temperature control, fiber scrambling and mechanical agitation, HZPF is designed to achieve a radial velocity precision < 3m/s, with a desire to obtain <1m/s for the brightest targets. This instrument will enable a study of the properties of low mass planets around M dwarfs; discover planets in the habitable zones around these stars, as well serve as an essential radial velocity confirmation tool for astrometric and transit detections around late M dwarfs. Radial velocity observation in the near-infrared (NIR) will also enable a search for close in planets around young active stars, complementing the search space enabled by upcoming high-contrast imaging instruments like GPI, SPHERE and PALM3K. Tests with a prototype Pathfinder instrument have already demonstrated the ability to recover radial velocities at 7-10 m/s precision from integrated sunlight and ~15-20 m/s precision on stellar observations at the HET. These tests have also demonstrated the ability to work in the NIR Y and J bands with an un-cooled instrument. We will also discuss lessons learned about calibration and performance from our tests and how they impact the overall design of the HZPF.

## 1. INTRODUCTION

With the discovery of over 450 extrasolar planets, considerable interest is now focused on finding and characterizing terrestrial-mass planets in habitable zones around their host stars. Such planets are extremely difficult to detect around F, G, and K stars, requiring either very high radial velocity precision (<<1 m/s), very high cadence, or space-based photometry to detect a transit. Ongoing space missions like ***KEPLER*** aim to discover the signature of transiting earth-like planets around solar type stars, but confirming such detections, even with the most precise radial velocity instruments currently available, is quite challenging. Many ongoing radial velocity programs are now beginning to focus on M dwarfs because the lower stellar luminosity (compared to the Sun) shifts the habitable zone (HZ), a region around a star where liquid surface water may exist on a planet, much closer to the star (Figure 1). The lower stellar mass of the M dwarfs, as well as the short orbital periods of HZ planets, increases the Doppler wobble caused by a terrestrial-mass planet. Planets around early M stars are already detectable with the current radial velocity precision obtained with high-resolution echelle spectrographs. Figure 1 also shows the expected radial velocity amplitude induced by various low mass planets around M stars. A total of 19 exoplanets in 14 planetary systems have, to date, been discovered around M stars, including the low mass exoplanets around GJ581, GL436, the saturn mass planets around GJ1148, and GL649. These observations suggest that, while hot Jupiters may be rare, lower mass planets do exist around M stars and may be rather common.

*suvrath@astro.psu.edu; phone 1 814 865-0261

Theoretical work based on core-accretion models and simulations also predicts that short period Neptune mass planets should be common around M stars[1]. Climate simulations of planets in the HZ around M stars[2] show that tidal locking does not necessarily lead to atmospheric collapse, and a surface pressure of 1-2 bars allows liquid water to exist. The habitability of terrestrial planets around M stars has been explored and studies[3,4] find that, despite their flares and low flux, M dwarfs are good candidates for hosting habitable planets. M dwarfs are the most numerous stars in the Galaxy, and an understanding of their planet population is essential to estimate $\eta_{earth}$, the fraction of earth-like planets in our Galaxy.

While many optical radial velocity surveys, which currently cover the 0.38-0.68μm range (or a subset of that), include late type targets (HARPS, M2K, CPS), nearly all of the M stars in these surveys are typically earlier in spectral type than M4 because late-type M stars are intrinsically faint in the optical: they emit most of their flux in the 0.9-1.8 μm wavelength region, the near infrared (NIR) Y (0.9-1.1 μm), J (1.1-1.4 μm) and H (1.45-1.8 μm) bands. These stars are also very desirable targets because the velocity amplitude of a terrestrial planet in the habitable zone is larger than for more massive stars. Unlike early type stars (A, B, early F) M stars **do** have sharp absorption lines and molecular features, which contain significant radial velocity information, making them suitable for exoplanet searches. Because the flux distribution from M stars peaks sharply in the NIR, a stable high-resolution NIR spectrograph capable of delivering high radial velocity precision can observe the nearest M dwarfs to examine their planet population. Lessons learned from our ongoing Pathfinder testbed effort[5] enable us to take the next steps to deliver 3 m/s or better radial velocity precision with the Habitable Zone Planet Finder (HZPF). The NIR spectral region is also rich in telluric absorption lines and OH emission lines. High spectral resolution can mitigate the effect of these atmospheric contaminants. The HZPF will be capable of a nominal resolution of R~50,000, which is expected to be sufficient to be able to mitigate these effects.

The RECONS team (*www.recons.org*) has determined that 72% of the 354 objects currently known to be within 10pc of

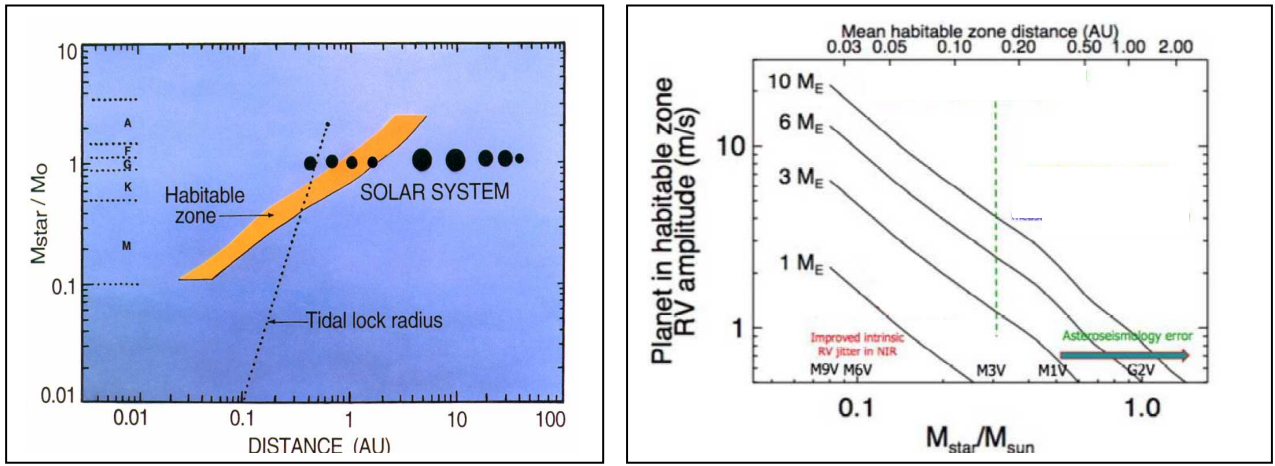

Figure 1. The Habitable Zone around main sequence stars[6] and the RV signal expected from planets in the Habitable Zone around M stars[7]

the Sun are M stars, with a significant number of these being low mass M stars suitable for a NIR surveys.

Ground based transit surveys, like MEarth are now beginning to target mid-late type M dwarfs to search for transiting super-Earths. These surveys **require** radial velocity observations to confirm transit detection, as well as to determine a mass for the transiting object Attempts to prove/disprove the claim of an astrometric planet detection around the M8V star vB10[8] with NIRSPEC on Keck did not entirely settled the question until the absence of any radial velocity period was demonstrated using an ammonia cell with CRIRES at the VLT[9]. The HZPF will play an important role in this emerging discovery regime, whereas such complementary activities would be impractical in the optical (for example, vB10 has a V magnitude of 17.3, but is much brighter in H at 9.2). Stellar activity, most frequently associated with young stars, complicates the detection of exoplanets and radial velocity observations of young stars in the optical and NIR are essential for discriminating between activity and the presence of planets. NIR spectroscopy can explore new regimes of planet formation, as well as complement existing observational programs. NIR radial velocity searches to find close-in planets, coupled with imaging to find the long-period planets, have the potential to address the relative importance of various formation channels in the early epoch of planet formation. Young planets tend to be more self-luminous, making potential detection with JWST and other space missions easier.

These compelling scientific pursuits, many of which require a stable high-resolution NIR spectrograph on a large telescope, are the primary motivators for the development of the Habitable Zone Planet Finder concept.

## 2. TARGETS FOR A SURVEY WITH THE HET

To ensure the feasibility of our scientific goals we have already begun to generate target lists for M stars in the

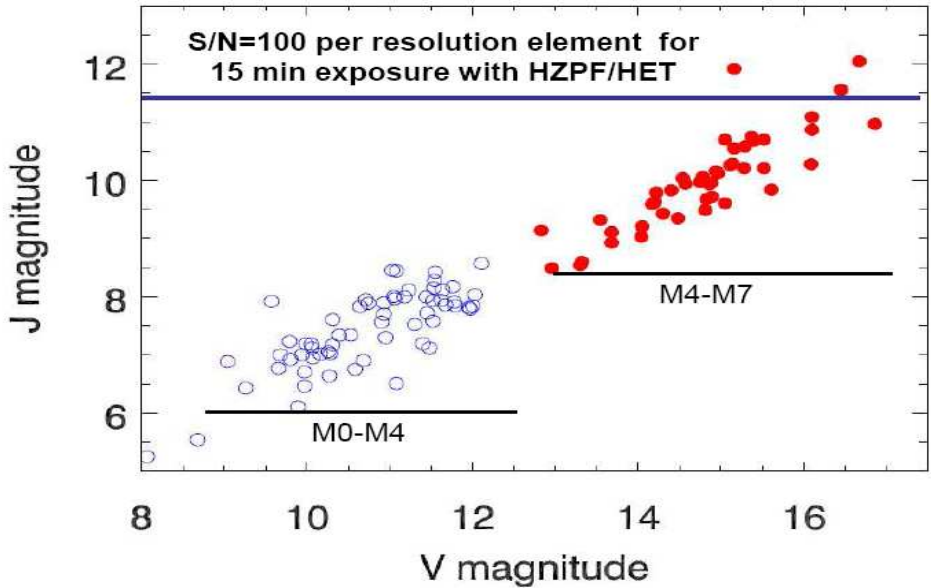

Figure 2. V and J magnitudes for stars in the current optical HET sample (open blue circles) and from late-type M dwarf survey for rotation velocity (filled red circles).

Mass range 0.08-0.4 $M_{sun}$, spanning a factor of 5 in mass and extending the parameter space of discovery to complement ongoing radial velocity surveys around 0.4-1.5 $M_{sun}$ stars. The majority of our targets will be within 20 pc. As part of an ongoing effort to measure the rotational velocity of possible targets for a high precision RV survey we have acquired observations of 56 main sequence stars of spectral type M4-M7 with the red orders of the High Resolution Spectrograph (HRS) on the HET[10] (Figure 2). When combining our work with known *vsini* measurements from the literature we find 138 M4-M9 stars known to have rotational velocities less than 10 km/s. Since the radial velocity information content available in the M star spectra degrades by a factor of ~3.5 as *vsini* rises, slower rotators are better targets for a radial velocity survey. The target selection will however also include some faster rotators to mitigate selection effects, especially at M7 and later spectral types since slow rotating stars are relatively uncommon in those regimes. We are continuing to refine our target sample with an ongoing survey at the HET and hope to have a well defined target sample of ~300 M4-M9 stars over the next few years.

## 3. THE NEED FOR A STABLE NIR SPECTROGRAPH

The main radial velocity techniques currently in use in the optical region are to use an Iodine ($I_2$) cell inserted into the stellar beam path, or to use a calibration fiber coupled to a Th-Ar emission lamp. The $I_2$ technique uses an existing very high resolution spectrum of the $I_2$ cell to model the spectrograph PSF variation. Precision radial velocities can only be recovered using the spectral region where $I_2$ has dense absorption lines (~0.5-0.62μm). The Th-Ar technique requires a stabilized spectrograph and fiber-feeding to mitigate seeing and image drifts on the slit, but the entire spectral bandpass of the instrument can, in principle, be used to derive the radial velocity. However it is in the NIR that the M dwarfs, emit most of their light. Thus a stable NIR spectrograph is an essential tool to discover terrestrial mass planets around these relatively unexplored stars. In the 0.85-1.7μm region that we wish to observe there are no molecular gas cells (like $I_2$) that have a dense set of sharp absorption lines over a significant bandwidth that one can use (though gas combinations are being explored.[11,12]

Table 1. Current, Planned and Proposed NIR Spectrographs capable of precision velocity measurements

| Instrument | Coverage | Simultaneous coverage | Resolution | Fiber Fed | Telescope | Status | Reference |
|---|---|---|---|---|---|---|---|
| NAHUAL | 0.9-2.4μm | 0.9-2.4μm | R=50,000 | Maybe | GTC 10m | proposed | Martin et al. 2005 |
| PHOENIX | 1-5μm | ~0.08μm | R=70,000 | No | Gemini 8m | available | Hinkle et al. 2000 |
| UPF | 1-1.75μm | 0.9-1.75μm | R=70,00 | Yes | UKIRT 4m | proposed | Jones 2009 |
| CRIRES | 1-5μm | ~0.05μm | R=100,000 | No | VLT 8m | available | Käufl et al. 2008 |
| NIRSPEC | 1-5.5μm | ~0.2μm | R=25,000 | No | Keck 10m | available | McLean et al. 1998 |
| GIANO | 0.9-2.5μm | 0.9-2.5μm | R=46,000 | No | TNG 4m | in progress | Oliva et al. 2006 |
| FIRST | 1.3-1.8μm | 1.3-1.8μm | R=50,000 | Yes | ARC 3.5m | proposed | Ge et al. 2006 |
| TEDI | 0.9-2.4μm | 0.9-2.4μm | R~2700 | Yes | Palomar | In progress | Edelstein et al. 2007 |
| WINERED | 0.9-1.35μm | 0.9-1.35μm | R~100,000 | No | Subaru 8m | In progress | Yasui et al. 2008 |
| CARMENES | 0.9-1.75μm | 0.9-1.75μm | R~80,000 | Yes | CalarAlto 4m | proposed | Quirrenbach 2010 |
| **HZPF** | **0.85-1.7μm** | **0.85-1.7μm** | **R~50,000** | **YES** | **HET 10m** | **proposed** | **this paper** |

We conclude that the velocity requirements to effectively detect low mass planets (~3m/s or better) are most readily achieved with an NIR fiber-fed instrument that has a second calibration fiber to track instrument drifts, enabling full use of the spectral information, and also not limiting observations to specific (gas cell) regions which may have high telluric contamination. In fact with such an instrument, our fiber-fed prototype Pathfinder, we have already demonstrated 7-10 m/s short term precision on sunlight using an IR array in the Y band, as well as obtained radial velocities on stars with the Hobby Eberly Telescope. **Table 1** lists existing and planned high resolution NIR spectrographs or instruments designed to measure RV. Most instruments listed either have very small simultaneous wavelength coverage, or are not fiber-fed (making them reliant on gas cells that are yet to be demonstrated), or cannot cover the information rich Y band. The proposed UPF for the UKIRT 4m telescope and HPZF have their heritage from the Gemini commissioned PRVS instrument study, in which Penn State was a partner. The PRVS design won the Gemini instrument competition though the final instrument was cancelled during contract negotiations by the Gemini Board. The HZPF on the HET will be one of the few (or only) fiber-fed instrument on a 10m class telescope, capable of high-resolution, high stability, and large simultaneous wavelength coverage.

## 4. THE HABITABLE ZONE PLANET FINDER

Most existing NIR high resolution spectroscopic instruments either have very small simultaneous wavelength coverage, or are not fiber-fed, or cannot cover the information rich Y band. The HZPF on the HET will be a cooled fiber-fed instrument on a large telescope capable of high-resolution, high stability, and large simultaneous wavelength coverage. Our HZPF optical and system design draws heavily on the lessons learned in the PRVS design study. We have explored various design options that meet the science requirements and the baseline approach we adopt is the well proven quasi-Littrow white pupil design with a monolithic off-axis parabolic collimator, a 200mm by 800mm 31.6 g/mm replicated mosaic grating from Newport RGL blazed at 75 degrees, a grism cross disperser with a 150 l/mm 5.4 deg blaze grating on it, and an f/2.3 refractive camera made with standard Schott and Ohara glasses. The white pupil design enables good control of scattered light, and the grism cross-disperser enables a compact configuration. Our low risk backup option (in case the grism cross disperser proves to be difficult to manufacture) is a 150 l/mm reflection grating from Newport. The beam size of 150mm enables all the refractive optics to be of reasonable size, while still enabling a 0.5 arc second slit to be sampled by 3 pixels on the H2RG (3.6 pixels on the CCD) to yield a resolution of R~50,000. Figure 3 shows the quantum efficiency of a deep-depletion CCD and the H2RG and the order format and the split over the CCD and H2RG although this may be revised as we continue our trade studies. A point worth highlighting is that the current split between the CCD and H2RG loses part of the Y band. Tests with our Pathfinder instrument prototype reveal that the Y band is the least affected by telluric contamination and rich in radial velocity information content. We plan to revisit the choice of the region split to take advantage of our region. The simultaneous use of a CCD enables the instrument to be fully tested while warm, as well as providing an independent source of RVs that are not affected by any of the possible systematics of NIR arrays. The monolithic collimator mirror, Echelle grating, fold mirror, and cross disperser grating will all be manufactured on Zerodur substrates to reduce temperature sensitivity, and will all be gold coated to increase efficiency. The well-understood white pupil design enables good control of the scattered light, as well as minimizing the size of the cross-disperser and camera apertures. RMS spot sizes from our preliminary optical design are all within 1 pixel for the H2RG detector and within 1.5 pixels for the CCD.

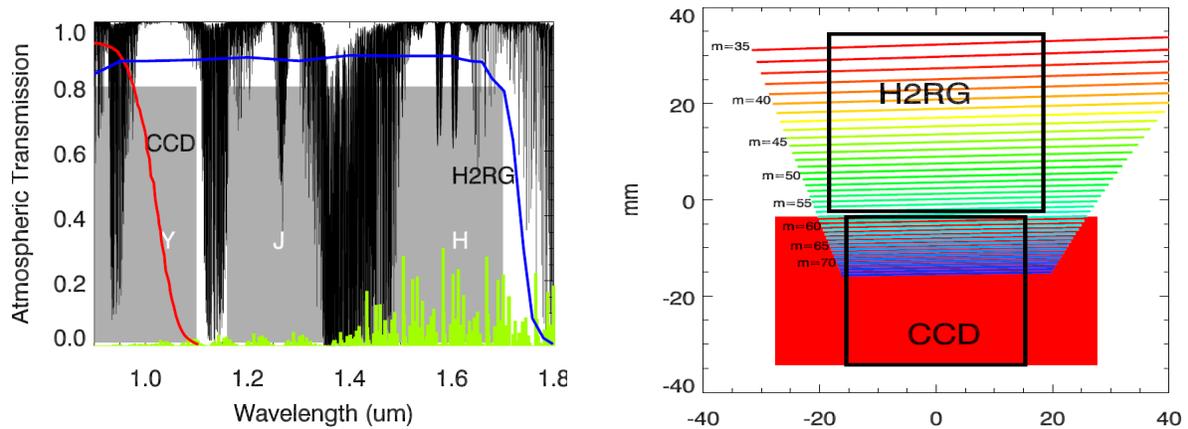

Figure 3. (LEFT) Atmospheric Transmission, OH emission (green lines), and the quantum efficiencies of the CCD (red line) and H2RG (blue line). (RIGHT) Format of the HZPF Spectral Orders. A filter cuts off in the blue at 0.85um to prevent order overlap from the cross-disperser.

Figure 4 shows the schematic layout of the HZPF coupled to the HET. The use of the fiber optics allows not only good scrambling of the input PSF, but also enables the spectrograph to be located in the stable sub-basement of the HET. The use of both an optical double scrambler and a mechanical agitator ensures that PSF and modal noise issues will be mitigated at the level necessary for high velocity precision, and the optical fiber slicer enables high efficiency. A simultaneous calibration fiber coupled to a lamp will be used to track the velocity drift of the spectrograph. Successful implementation of the calibration technique we plan to utilize requires high thermal stability in the spectrograph. HARPS has pointed the way to achieving high stability by placing the instrument in a vacuum in a stable thermal environment. HZPF will also be in a vacuum vessel, but has the additional complication of needing to be cooled so that thermal radiation is not a significant contributor to the system background. As part of the PRVS detailed study, we determined that the spectrograph components and enclosing structure need to be at < 200 Kelvin. This temperature will yields a thermal background per pixel in the H2RG of 0.1-0.2e/s/px. Thus, while not fully cryogenic, HZPF will be structurally more similar to cryogenic infrared instruments than optical instruments. Our baseline design approach consists of a vacuum chamber which is supported with optical bench style vibration damping supports. All the spectrograph optical components plus the detector are mounted on an optical support structure constructed of aluminum hollow rectangular beams-weldments or light-weighted aluminum plates so as to have the necessary stiffness. The optical support structure will be supported at three points by flexures from the bottom of a radiation shield which is itself supported by insulated flexures from the bottom of the vacuum chamber. The top of the radiation shield and vacuum chamber will be removable for access to the optical support structure via pneumatic lifts. Figure 4 shows the schematic layout concept for the spectrograph, and Figure 5 the support structure, radiation shield, and vacuum vessel. The optical support structure and radiation shield will be maintained at ~195 K by using resistive heating elements on the optical support structure and a vibration isolated CT1050 closed-cycle cooler. Using a Lakeshore 340 temperature controller we expect temperature stability to better than 0.03 K. Active temperature stabilization of HZPF is critical for reducing the instrument drift and achieving high precision radial velocities. Small temperature gradients between different parts of the instrument are acceptable as long as the gradients are stable. A second closed-cycle cooler will be used for the detector. Fiber inputs to the vacuum chamber are baselined as optical feed-throughs. By the time of a possible HZPF commissioning the Hobby Eberly Telescope Dark Energy Experiment (HETDEX) instruments will be operational and the HZPF provides a perfect bright time complement to the dark-time HETDEX project.

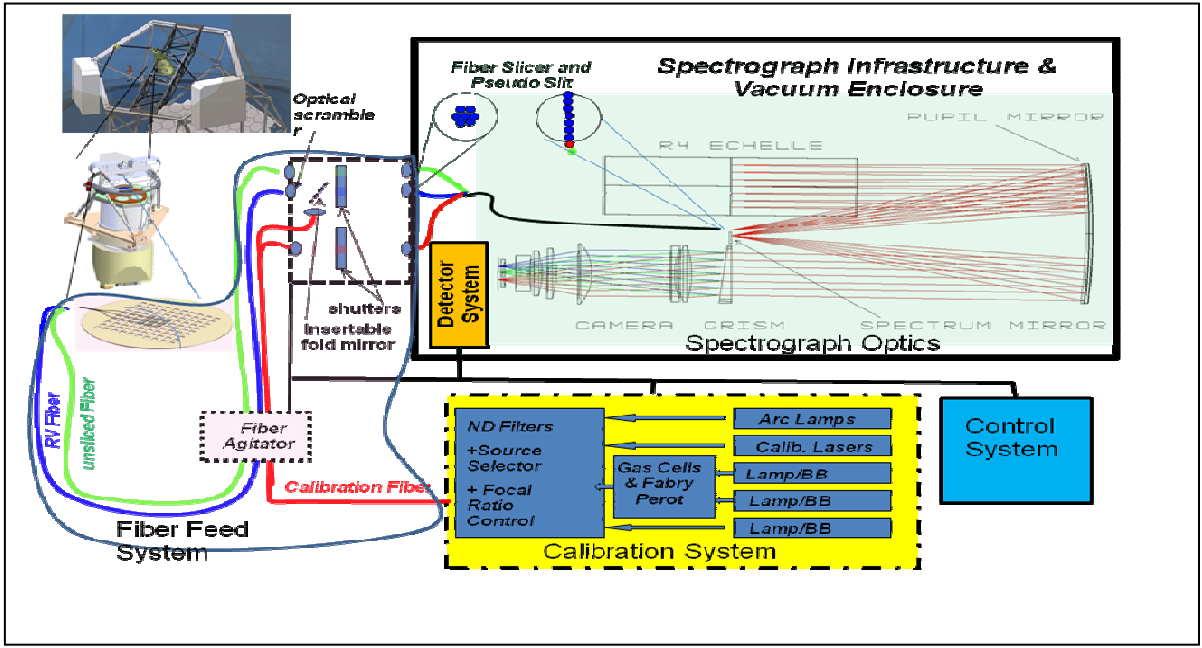

Figure 4. Schematic Drawing of the HZPF Instrument with the major subsystems shown

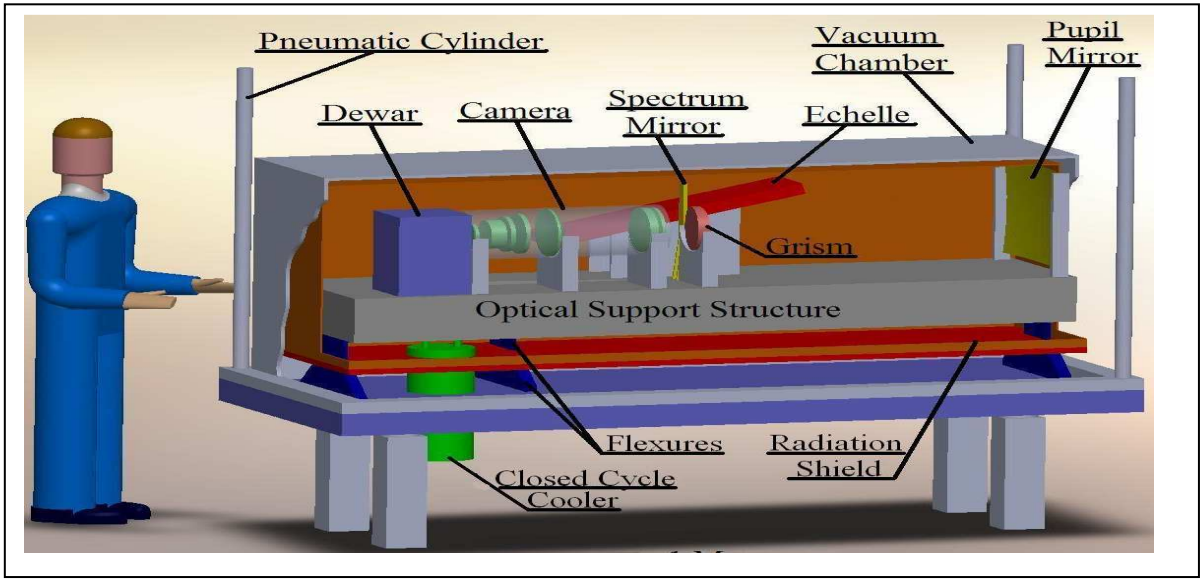

Figure 5. Cut out view of HZPF design concept showing the spectrograph optics, vacuum vessel, support structures and radiation shields.

# 5. THE PATHFINDER TESTBED: CURRENT STATUS & LESSONS FOR HZPF

The current Pathfinder testbed is a fiber-fed cross-dispersed uncooled Echelle spectrograph brass-board instrument equipped with a 1kX1k Hawaii-1 NIR array that is sensitive to 2.5μm. Pathfinder was been built and designed to allow us to probe the limits of precision radial velocity in the NIR, test calibration schemes, and significantly retire risk for future instruments like the PRVS and HZPF. After achieving 7-10m/s precision in lab tests with sunlight[5], Pathfinder has been redesigned and moved to the Hobby Eberly Telescope in the first quarter of 2010. For the upgraded Pathfinder the collimator is a simple 609 mm gold coated parabolic mirror. The primary dispersing grating is a 31.6g/mm 71.5º echelle that is operated at an in-plane, but non-Littrow angle. The in-plane design virtually eliminates variable slit tilt effects caused by an out-of-plane angle, enabling the exploration of fiber slicers and long slits in this configuration. This grating provides a resolving power of R~50,000; with a 100um slit the resolution element is sampled by 4.4 pixels. The diffracted beam is cross-dispersed by a 150 line/mm $\Theta_{blaze}$ = 5.4 degree gold coated 135 x 165mm grating on Zerodur substrate. The Camera in this instrument is, like the collimator, an aluminum coated on-axis commercial parabolic mirror made of pyrex from Edmund Optics. We use a weak lens near the focus to correct most of the coma to well within one pixel at the edge of the detector field of view. The camera, with an effective 800 mm focal length, is folded so as to allow the dewar containing the detector array to be mounted vertically for maximum cooling efficiency. The order coverage on the Hawaii 1K array is currently 40% of a free spectral range (FSR) at 1 micron (Y band) decreasing to 32% at 1.35 microns (J band). While our NIR detector is sensitive out to the K band, our preference is to work in the Y band (~0.98-1.1μm) where M dwarfs have many sharp lines, telluric absorption is lowest, and OH emission also small. A significant development is the use of thermal blocking filters and absorptive glass combinations to suppress the 1.35-3μm thermal background, enabling us to work in the NIR Y and J bands with an un-cooled instrument. Figure 6 shows the Pathfinder set up at the Hobby Eberly Telescope. Figure 7 shows the ~8-9 orders covered in the Y band, with the spectrum of Tau Boo and a simultaneous calibration fiber obtained with the Pathfinder instrument at the HET in April 2010.

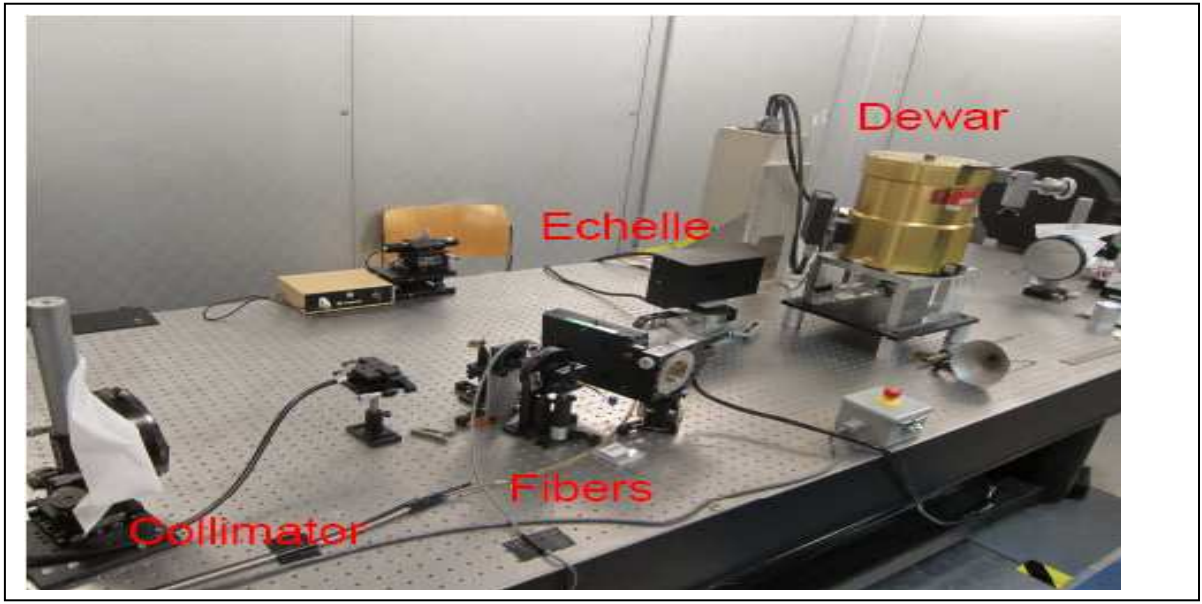

Figure 6. The Pathfinder Instrument testbed set up at the HET spectrograph basement. The instrument is fiber coupled to the telecope focal plane.

## 5.1 Major Technical Developments in the Pathfinder Project

Our continuing efforts to develop techniques to achieve precise radial velocities in the NIR has resulted in a number of innovations in the Pathfinder project that make a warm NIR spectrograph and precise wavelength calibration possible in this wavelength regime

*Wavelength Calibration Sources Suitable for the NIR:* Our science goal of eventually achieving a radial velocity precision of <3 m/s with HZPF requires that the instrumental drift be measurable to <1 m/s to ensure that it is a small fraction of the total error budget. This makes some form of simultaneous calibration essential. There are as yet no suitable gas cells like $I_2$ covering the Y, J and H band in the NIR, but the fiber-fed design of the Pathfinder allows a calibration fiber to track the instrument drift during an object exposure. The major requirements on the calibration source are that there be sufficient number of bright stable emission lines in each echelle order to achieve <1 m/s by tracking drifts of the spectral response function (PSF). Inexpensive hollow cathode lamps with stable heavy elements, like Thorium ($^{232}$Th), have sharp emission lines, making them suitable for such applications. Such lamps also have a filler gas, usually (Ar) or Neon (Ne), which can be very bright, but are not stable enough for our use since pressure shifts make them sensitive to environmental conditions in the lamp. Th-Ar emission lines span the UV-NIR regions, making them a very useful and convenient source of wavelength calibration. They are used with the fiber-fed HARPS spectrograph to achieve radial velocity precisions better than 1 m/s and discover super-Earth planets. In the NIR however the Ar lines are very bright, completely dominating the flux output of the lamp. There are many suitable Thorium lines for use in wavelength calibration, but unlike the situation in the optical the flux output of these lines can be smaller by a factor of $10^2$ -$10^4$ compared to the bright Argon lines. The Argon lines are unsuitable for use for precision radial velocity measurements at the few m/s. In the NIR long integrations are therefore needed to have sufficient intensity in the Th lines to allow wavelength calibration, during which the Argon lines would cause some scattered light within any spectrograph.

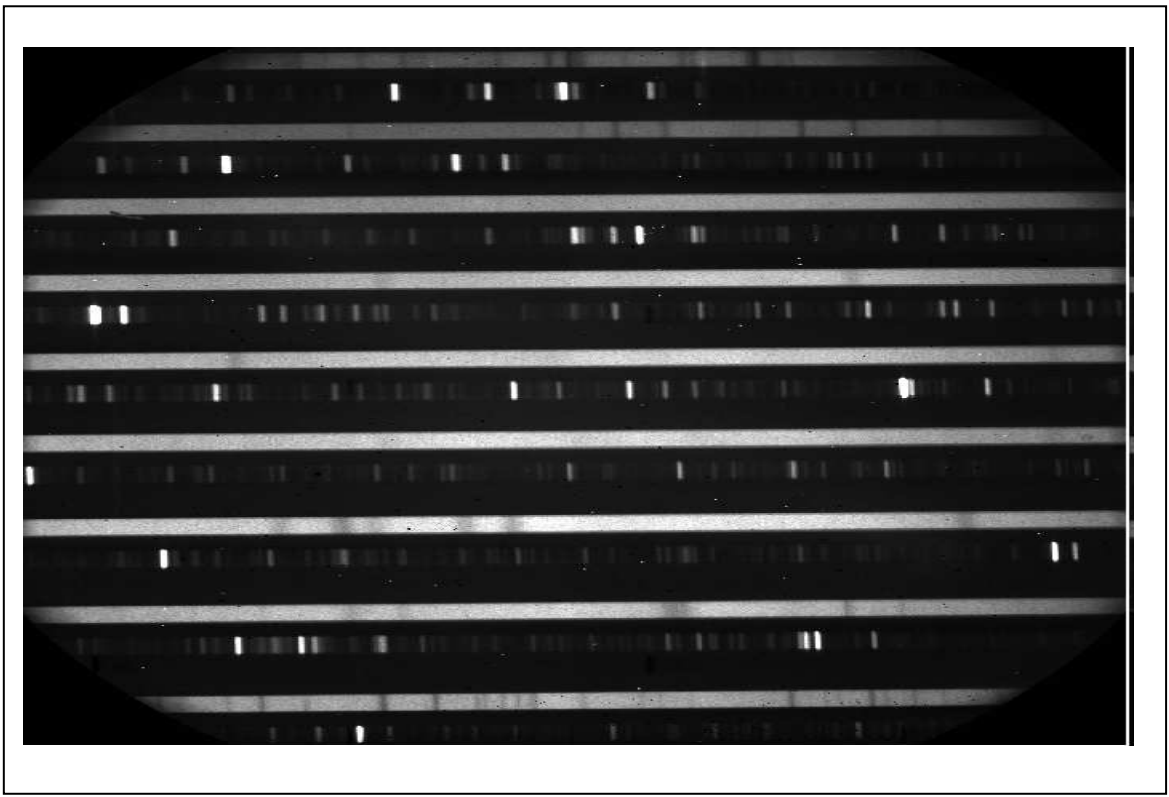

Figure 7. Raw frame of the spectra obtained with Pathfinder at the HET with Tau Boo and a Uranium-Neon Calibration lamp.

These bright lines from filler gases are useful for moderate precision wavelength calibration, but a hindrance for instruments such as Pathfinder and HZPF that require very high precision calibration. Our measurements with Pathfinder show that lamps with Neon as a filler gas have fewer bright lines in the NIR Y band than Argon, causing less scattered light and pixel saturation. Our experiments with Uranium (U) lamps convincingly show that U has a large number of emission lines in the Y band, making it an attractive alternative to Thorium. Uranium ($^{238}$U) fulfills many of the requirements needed of an element in a hollow cathode lamp for wavelength calibration: it is a heavy element (heavier than Thorium), has zero nuclear spin, and has a long half-life. Its only disadvantage is the presence of a very small amount (~0.7%) of $^{235}$U in all naturally occurring Uranium, which may lead to very subtle isotope shifts between different lamps. A U/Ne hollow cathode lamp is now the calibration lamp we use for Pathfinder to enable accurate wavelength calibration as well as simultaneous monitoring of the instrument drift. A detailed line list for U with accurate wavenumbers is now in preparation by our team is collaboration with NIST. We will extend our analysis to the J and H band to determine if U is better than Th in these bands as well. The development and careful consideration of such calibration techniques is essential for the HZPF to meet its precision goals.

***Blocking of the Thermal NIR for an un-cooled NIR spectrograph:*** Many groups over the years have attempted to use a 2.5μm Hawaii detector array in a warm instrument to observe in the Y and J band with short pass interference filters and discovered that observing even moderately faint targets is impossible because the detector is usually swamped by the thermal background with moderate exposures, even with the use of cutoff-filters blocking to 1.4-2.5μm to eliminate H and K. This is because a 2.5μm Hawaii array is actually weakly sensitive to light in the 2.5-2.8μm region where the background is **significantly** higher as well. The obvious solution is to cool down the spectrograph to cryogenic temperatures. However this is expensive and complicated, and certainly not justifiable for a testbed like Pathfinder. *We have significantly mitigated the thermal background problem for observations in the Y band using a PK50 glass filter to*

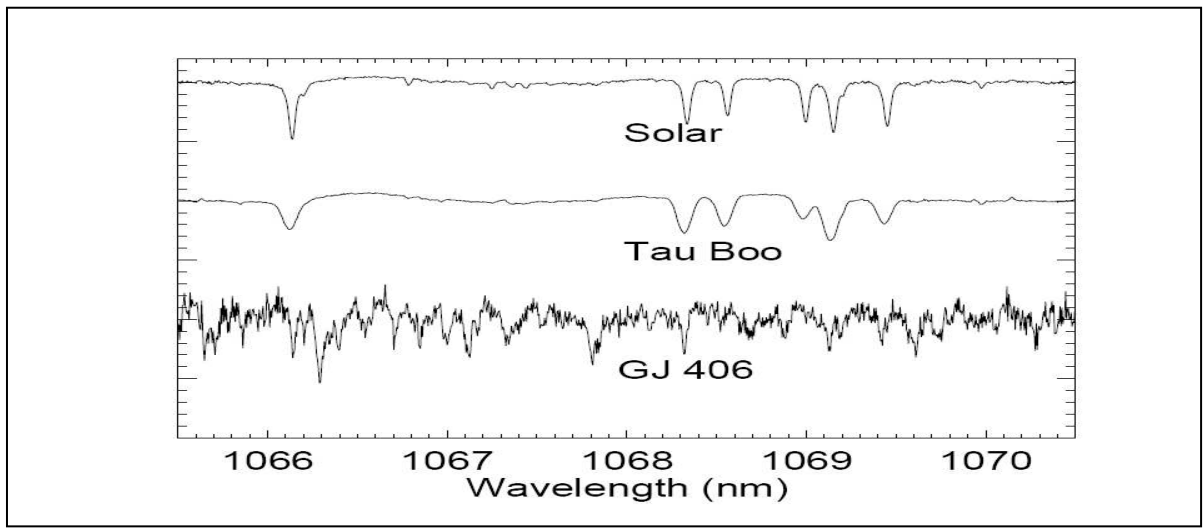

Figure 8. Spectra obtained with the Pathfinder Instrument at the Hobby Eberley Telescope

*absorb thermal radiation > 2.7μm coupled with a custom dual magnetron reactive sputtering coated filter and a commercially available Spectrogon-1300 short pass filter to further suppress the thermal background at $10^5$ level from 1.3-3μm.* Instead of the whole instrument being kept cold only the filters and the detector are kept at liquid nitrogen temperatures of ~77-80K. Our current detector is a Hawaii-1 science grade array that was originally procured for the red arm of the HET Medium Resolution Spectrograph (MRS) instrument, called JCAM. The system was provided nearly a decade ago by IR Labs in Tucson Arizona and uses a SDSU Gen II controller. Inside the LN2 dewar radiation shield we have an Aluminum filter holder mounted to the same cooling block as the detector that keeps the blocking filters and the PK50 at cryogenic temperatures. This temperature is monitored and is typically 80 K. The cold stop is the short pass filter inside the dewar window. Thus, the detectors views ~0.1 steradian of solid angle of room temperature and any spectral leakage in the filter will show up as background. The 12.5mm PK50 glass, however, ensures that virtually no radiation longer than ~2.7μm is incident on the detector. With this setup the combined thermal background and dark counts from the detector *are < 4 e/px/sec*, easily enabling 10 minute observations of reference stars and M dwarfs with Pathfinder on the HET. A new NIR array that is sensitive only to 1.7μm (current one is sensitive to 2.5 μm) will help mitigate the thermal background and read noise even further. The acquisition of such a NIR array (Hawaii-2RG) and the SIDECAR ASIC controller for a Pathfinder upgrade has recently been funded by NSF through the ATI program. This

## 5.2 Pathfinder On Sky at the HET

The Pathfinder testbed was originally designed to retire the risks of the Gemini PRVS proposal. With further development, and testing, both in the lab, and on-sky at the HET, we have demonstrated that precision radial velocities are indeed possible in the NIR. The Pathfinder instrument testbed was set up at the HET in the first quarter of 2010 and has been used in two observing runs so far. Spectra acquired on diffuse solar light, Tau Boo, and GJ406 from one of these runs are shown in Figure 8. Pathfinder is already achieving 15-20m/s radial velocity precision on many stable stars, and analysis of the data is still ongoing. A detailed description of the data analysis and observations, and results is beyond the scope of this paper, since the focus of this work is the HZPF. The results of the Pathfinder observations will be published separately, but are mentioned here in the context of demonstrating the HZPF concept. Pathfinder, being a stable fiber-fed testbed, is also an excellent resource for the testing of laser- frequency combs that are now being developed in the NIR. The development of new calibration sources, and techniques to mitigate the thermal background will also fold into the design of the HZPF instrument. The continuing development of the Pathfinder Testbed mitigates many of the risks and unknows involved with precision measurements in a new wavelength regime, and leading to the development of the facility-class Habitable zone planet finder on the 10-m Hobby Eberley Telescope.

We are grateful to Hugh Jones for useful discussions about HZPF. We acknowledge the significant heritage from the Gemini PRVS effort. We acknowledge support from NASA through the NAI and Origins grant NNX09AB34G and the NSF ATI program with NSF grant 1006676. This work was partially supported by funding from the Center for Exoplanets and Habitable Worlds. The Center for Exoplanets and Habitable Worlds is supported by the Pennsylvania State University, the Eberly College of Science and the Pennsylvania Space Grant Consortium. Part of this work is based on observations with the Hobby Eberley Telescope. The Hobby-Eberly Telescope (HET) is a joint project of the University of Texas at Austin, the Pennsylvania State University, Stanford University, Ludwig-Maximilians-Universität München, and Georg-August-Universität Göttingen. The HET is named in honor of its principal benefactors, William P. Hobby and Robert E. Eberly